\pdfoutput=1
\documentclass[aps,prl,showpacs,twocolumn,superscriptaddress]{revtex4-1}

\usepackage{amsmath}
\usepackage{amssymb}
\usepackage{graphicx}
\usepackage{hyperref}
\usepackage[arrow, matrix, curve]{xy}
\usepackage{bm}
\usepackage{yhmath}
\usepackage{color}

\newcommand{\bra}[1]{\langle #1|}
\newcommand{\ket}[1]{|#1\rangle}

\newcommand{\braket}[2]{\langle #1|#2\rangle}
\newcommand{\matrixbraket}[3]{\langle #1|#2|#3\rangle}
\newcommand{\scalar}[2]{#1 \cdot  #2}
\newcommand{\disorder}[1]{[#1]_\text{av}}

\newcommand{\opf}[1]{\hat{#1}}
\newcommand{\maf}[1]{#1}
\newcommand{\img}{\mathsf{i}}
\newcommand\diff{\mathrm{d}}
\renewcommand{\vec}[1]{\mathbf{#1}}

\DeclareMathOperator\Real{Re}

\usepackage[usenames,dvipsnames]{xcolor}
\hypersetup{colorlinks=true, linkcolor=BrickRed, urlcolor=blue!50!black, citecolor=blue!50!black}

\usepackage[normalem]{ulem}

\makeatletter
\newcommand\hide@visible[1]{%
  \bgroup\fboxsep=.3ex\colorbox{Gray}{begin hide}%
  #1\colorbox{Gray}{end hide}\egroup%
}
\newcommand\hide@hidden[1]{%
  \bgroup\fboxsep=.3ex\colorbox{Gray}{hidden text}%
}
\newcommand\hide@invisible[1]{}
\newcommand\makevisible{\let\hide\hide@visible}
\newcommand\makehidden{\let\hide\hide@hidden}
\newcommand\makeinvisible{\let\hide\hide@invisible}
\makeatother
\makehidden

\begin{document}


\title{Nonlinear Response in the Driven Lattice Lorentz Gas}



\author{Sebastian Leitmann}

\author{Thomas Franosch}
\affiliation{Institut f\"ur Theoretische Physik, Leopold-Franzens-Universit\"at Innsbruck, Technikerstra{\ss}e~25/2, A-6020 Innsbruck, Austria}
\email[]{thomas.franosch@uibk.ac.at}
\affiliation{Institut f\"ur Theoretische Physik, Friedrich-Alexander-Universit\"at Erlangen-N\"urnberg, Staudtstra{\ss}e~7, 91058 Erlangen, Germany}



\date{\today}

\begin{abstract}
We determine the nonlinear time-dependent response of a tracer on a lattice with randomly distributed hard obstacles as a force is switched on. 
The calculation is exact to first order in the obstacle density and holds for arbitrarily large forces. 
Whereas, on the impurity-free lattice, the nonlinear drift velocity in the stationary state is analytic in the driving force, interactions with impurities introduce logarithmic contributions beyond the linear regime.
The long-time decay of the velocity toward the steady state is exponentially fast for any finite value of the force, in striking contrast to the power-law relaxation predicted within linear response. 
We discuss the range of validity of our analytic results by comparison to stochastic simulations.
\end{abstract}



\maketitle


\paragraph{Introduction.---}
One of the principal strategies to probe material properties is to apply an external stimulus and monitor the system's response. 
Within statistical physics the fundamental link between the deterministic response of a system and the correlation functions 
of intrinsic fluctuations is provided by the celebrated fluctuation-dissipation theorem (FDT), tracing back to Nyquist~\cite{Nyquist:1928}, 
Onsager~\cite{Onsager:1931}, and later proved by Callen and Welton~\cite{CallenWelton:1951}. The framework applies whenever 
the unperturbed system is in thermal equilibrium and the forces are sufficiently small such that the response is linear.

It came as a big surprise that the equilibrium correlation functions yielding the transport coefficients via Green-Kubo relations~\cite{KuboTodaBook:1991, DorfmanBook:1999, EvansMorrissBook:2008, ResiboisdeLennerBook:1977, McLennanBook:1988} 
display power-law tails rather than an exponential decay. First found in computer simulations of the velocity autocorrelation function (VACF) for hard 
spheres~\cite{AlderWainwright:1967, AlderWainwright:1970} these persistent correlations have been derived rigorously for dilute gases~\cite{DorfmanCohen:1970}
by systematically going beyond the Boltzmann equation. Then repeated collisions with the same particle yield a nonanalytic dependence of the diffusion 
coefficients on density, frequency, and wave number. 
A similar exact low-density expansion for the related Lorentz model~\cite{vanLeeuwenWeijland:1967, vanLeeuwenWeijland:1968}, 
where a tracer ballistically explores a random array of fixed scatterers, again reveals long-time tails in the VACF $Z(t)$~\cite{ErnstWeyland:1971}.
For a planar lattice version of the Lorentz model, where a single particle performs a nearest-neighbor hopping dynamics and a certain fraction of sites is blocked by frozen hard obstacles, 
an exact solution for $Z(t)$ for all times to first order in the density of the obstacles~\cite{Nieuwenhuizen:1986, Nieuwenhuizen:1987:1, Nieuwenhuizen:1987:2} 
has been elaborated and confirmed by computer simulations~\cite{Frenkel:1987}.
The emergence of persistent memory~\cite{AlderWainwright:1967, AlderWainwright:1970, DorfmanCohen:1970, vanLeeuwenWeijland:1967, vanLeeuwenWeijland:1968, ErnstWeyland:1971, Nieuwenhuizen:1986, Nieuwenhuizen:1987:1, Nieuwenhuizen:1987:2, Frenkel:1987, KlagesBook:2007, GaspardBook:1998, Franosch:2011, Lukic:2005, FranoschHoefling:2010, Jeney:2008, Hoefling:2007, Hoefling:2013, KuboTodaBook:1991, DorfmanBook:1999, EvansMorrissBook:2008, ResiboisdeLennerBook:1977, McLennanBook:1988} is believed to be a generic feature of complex transport even beyond the exactly solvable limiting cases and constitutes one of the pillars of our current understanding of dynamics. 

As a consequence of the fluctuation-dissipation theorem (FDT) these slow power laws manifest themselves in the linear response with respect to driving.
However, in soft matter systems, already small forces have a strong effect on the dynamics and a plethora of new phenomena associated to the inherently nonlinear behavior emerges.
In particular, nonanalytic behavior has been demonstrated for a sheared fluid~\cite{ErnstCichocki:1978} and the thermostatted driven Lorentz gas~\cite{PanjaDorfman:2000}.
Although there have been significant recent advances for linear response around nonequilibrium states~\cite{BaiesiMaes:2009:PRL, Seifert:2012} and for steady states systems far from equilibrium via the fluctuation relations~\cite{Marconi:2008, EvansSearles:2008}, the time-dependent approach toward the steady state is in general unknown. 
\paragraph{Summary of main result.---}
We provide an explicit expression for the drift velocity $v(t)$ of a tracer on a planar square lattice Lorentz model to a step force $F\vartheta(t)$ of arbitrarily strong strength exact to first order in the obstacle density.
For sufficiently small driving the FDT already provides the connection of the drift $v(t)$ to the VACF $Z(t)$ in equilibrium by
\begin{equation}
 v(t) = \frac{F}{k_B T} \int_0^t \diff t' Z(t'), \qquad t>0.
\end{equation}
Without obstacles the VACF decays instantaneously, $Z(t) = D_0 \delta(t-0^+)$ implying a step response $v(t) = \vartheta(t) F D_0/k_B T$, where $D_0$ denotes the bare diffusion coefficient. Inserting obstacles suppresses the long-time diffusion coefficient $D = \int_0^\infty \diff t' Z(t')$ by introducing negative contributions to the VACF. In fact, persistent anticorrelation 
manifested in a long-time power-law decay $-A t^{-2}, A>0$, already emerges to first order in the obstacle density~\cite{Nieuwenhuizen:1986}. 
The long-time tail in $Z(t)$ suggests that $v(t)$ should display a slow algebraic approach $\sim t^{-1}$ toward the steady-state drift velocity $v(t\to \infty)$. However, our results show that for any finite driving the terminal velocity is reached exponentially fast in qualitative disagreement to the FDT. Thus, the range of validity of the FDT with respect to this feature shrinks to infinitesimally small forces. Equivalently, the nonanalytic frequency dependence of the response function is cut off at nonzero frequency and the crossover frequency is determined by the force. The terminal velocity depends nonmonotonically on the driving and displays a novel nonanalytic behavior. 

\paragraph{The model.---}
We consider  a finite  square lattice  $\Lambda  =\{ (x,y) \in \mathbb{Z} \times \mathbb{Z} : x,y \in [-L/2, L/2[\, \}$ of linear size $L\in 2 \mathbb{N}$, consisting of  
$N= L^2$  sites $\vec{r} = (x,y) \in \Lambda$. 
A particle performs a random walk on the lattice and the time evolution is governed by a master equation.
It is convenient to exploit the analogy of the master equation to a Schr\"odinger equation. 
Hence we consider the Hilbert space of lattice functions  $ \Lambda \to \mathbb{C}$  spanned by the orthonormal basis  of position kets $\ket{\vec{r}}$.
Then the site-occupation probability density $\ket{p(t)}$ obeys the  master equation $\partial_t \ket{p(t)}  = \opf{H} \ket{p(t)}$ where the (non-Hermitian) Hamiltonian $\opf{H}$ generates the dynamics. Inserting the position basis, the following form is obtained
\begin{align} \label{masterequationrp}
\partial_t \braket{\vec{r}}{p(t)} = \sum_{\vec{r}' \in \Lambda} \matrixbraket{\vec{r}}{\opf{H}}{\vec{r}'}\braket{\vec{r}'}{p(t)}.
\end{align}
Thus $\braket{\vec{r}}{p(t)}$ is the probability to find the walker at site $\vec{r}$ at time $t$ and $\matrixbraket{\vec{r}}{\opf{H}}{\vec{r}'}$ corresponds to the transition rate from $\vec{r}'$ to $\vec{r}$. We impose  periodic boundary conditions and consider  nearest-neighbor jumps $\mathcal{N} = \{\pm \vec{e}_x, \pm \vec{e}_y \}$ only. 
The  force  $F$ on the tracer (from now on measured in units of $k_B T / \text{lattice constant} $) introduces a bias  in the   
 corresponding transition probabilities $W(\pm \vec{e}_x)$ and $W(\pm \vec{e}_y)$.
Detailed balance $W(\vec{e}_x)/W(-\vec{e}_x) = e^F$ and $W(\vec{e}_y)/W(-\vec{e}_y)=1$, suggests
$W(\pm \vec{e}_x) = e^{\pm {F/2}} /(e^{F/2} + e^{-{F/2}} + 2)$
 and $W(\pm \vec{e}_y) = 1/(e^{F/2} + e^{-{F/2}} + 2)$ in accordance with Ref.~\cite{HausKehr:1987}. Here the rates add to unity, which fixes the unit of time. The Hamiltonian without impurities $\opf{H}_0$ then assumes the form
\begin{align}
\opf{H}_0 = \sum_{\vec{r} \in \Lambda}\big[ - \ket{\vec{r}}\bra{\vec{r}} + \sum_{\vec{d} \in \mathcal{N}} W(\vec{d})\ket{\vec{r}}\bra{\vec{r} - \vec{d}}\big].
 \end{align}
The dynamics in the presence of  randomly distributed obstacles on the lattice is generated by  the  Hamiltonian $\opf{H} = \opf{H}_0 + \opf{V}$, where 
$\opf{V}$ cancels the transitions to and from the impurity.  Furthermore we allow particles to start at an impurity site which then remain immobile. 
Explicitly for a single impurity at site $\vec{s}_1$, we write $\opf{V} = \opf{v}(\vec{s}_1) \equiv \opf{v}_1$ and the only nonvanishing matrix elements are $\matrixbraket{\vec{s}_1 }{ \opf{v}_1}{ \vec{s}_1} = 1$, $\matrixbraket{\vec{s}_1}{ \opf{v}_1 }{ \vec{s}_1 -\vec{d} } = - W(\vec{d})$, $\matrixbraket{\vec{s}_1 -\vec{d} }{ \opf{v}_1 }{ \vec{s}_1  } = -W(-\vec{d})$, $\matrixbraket{\vec{s}_1-\vec{d}}{ \opf{v}_1 }{ \vec{s}_1 -\vec{d} } = W(\vec{d})$. 
For $N_\text{i}$ impurities, $\opf{V}$ will be defined by $\opf{V} = \sum_{i=1}^{N_\text{i}} \opf{v}_i$. Strictly speaking this definition does not properly account for impurities on neighboring lattice sites, yet we shall be interested only in the lowest order in the density  $n = N_\text{i}/N$ in the limit of large lattices $N \to \infty$, where such configurations do not occur. Without driving, $F = 0$, an equilibrium state is provided by $\braket{\vec{r}}{p_\text{eq}} = 1/N$ invariant under time evolution by the choice of the dynamic rules.  

We shall assume that the force is switched on at time $t=0$ such that the thermal equilibrium state  $\ket{p(t=0) } = \ket{p_\text{eq} }$ evolves toward a new steady state $\ket{p(t\to \infty)}$.   

\paragraph{Solution strategy.---}
The time-dependent probability $\ket{p(t)} = \opf{U}(t) \ket{p_\text{eq}}$ for $t \geq 0$ is expressed in terms of the time-evolution operator $\opf{U}(t) = \exp(\opf{H}t)$ fulfilling the equation of motion
\begin{align} \label{diffeqtimeevolution}
\partial_t \opf{U}(t) = \opf{H} \opf{U}(t) ,\qquad \opf{U}(0) = 1.
\end{align}
Thus, the matrix elements $\matrixbraket{\vec{r}}{\opf{U}(t)}{\vec{r}'}$ represent the conditional probabilities for the walker to be at site $\vec{r}$ at time $t$ provided it started at $\vec{r}'$ at $t = 0$.
The goal is the moment-generating function $F(\vec{k},t)$ of the displacements $\vec{r}-\vec{r}'$,
\begin{align} \label{eq:momgentime}
 F(\vec{k},t) = \sum_{\vec{r}, \vec{r}' \in \Lambda} \exp[-\img \scalar{\vec{k}}{(\vec{r}-\vec{r}')}]\matrixbraket{\vec{r}}{\disorder{\opf{U}(t)}}{\vec{r}'} \braket{\vec{r}'}{p_\text{eq}}.
\end{align}
Here, $\disorder{\dotso}$ indicates an average over the disorder, and we already observed that $\braket{\vec{r}'}{p_\text{eq}}=1/N$ is site independent. 
For the impurity-free case $\opf{V} = 0$, the Hamiltonian $\opf{H} = \opf{H}_0$ is translationally invariant.
This property becomes manifest in the plane wave basis
\begin{align}
\ket{\vec{k}} = \frac{1}{\sqrt{N}} \sum_{\vec{r} \in \Lambda} \exp(\img \scalar{\vec{k}}{\vec{r}}) \ket{\vec{r}}, 
\end{align}
where $\vec{k} = (k_x, k_y) \in \Lambda^* = \{(2\pi x / L,2\pi y / L) : (x,y) \in \Lambda \}$, and  $\scalar{\vec{k}}{\vec{r}} = k_x x + k_y y$.
Every translationally invariant operator is diagonal in this basis, in particular $\opf{H}_0 \ket{\vec{k}} = \epsilon(\vec{k}) \ket{\vec{k}}$ with eigenvalue 
$\epsilon(\vec{k}) = - \sum_{\vec{d} \in \mathcal{N}}[(1 - \cos(\scalar{\vec{k}}{\vec{d}}))W(\vec{d}) + \img \sin(\scalar{\vec{k}}{\vec{d}}) W(\vec{d})]$.

The general case $\opf{V} \neq 0$, is conveniently expressed relying on the scattering formalism borrowed from quantum mechanics~\cite{BallentineBook:2003}.
We define the propagator (Green function) by the Laplace transform of the time-evolution operator
\begin{align}
\opf{G}(E) =  \int_0^\infty \diff t \exp(-Et) \opf{U}(t) = (E - \opf{H})^{-1}.
\end{align}
Comparison with Eq. \eqref{eq:momgentime} reveals that $\matrixbraket{\vec{k}}{\disorder{\opf{G}}}{\vec{k}} = \int_0^\infty \diff t \exp(-E t) F(\vec{k}, t)$.
Since the disorder-averaged propagator is again translationally invariant, only the matrix elements $\disorder{\maf{G}}(\vec{k}) = \matrixbraket{\vec{k}}{\disorder{\opf{G}}}{\vec{k}}$ are nonvanishing and conveniently expressed as 
\begin{align} \label{eq:self_energy_opf}
\disorder{\maf{G}}(\vec{k}) = \frac{1}{\maf{G}_0(\vec{k})^{-1} - \maf{\Sigma}(\vec{k})}.
\end{align}
Here the self-energy $\maf{\Sigma}(\vec{k})$ accounts for the interaction of the tracer with the frozen disorder and the free propagator $\maf{G}_0(\vec{k}) = \matrixbraket{\vec{k}}{\opf{G}_0}{\vec{k}} =  [E - \epsilon(\vec{k})]^{-1}$ drives the dynamics in the absence of obstacles. 

Similarly, the scattering operator $\opf{T} = \opf{V} + \opf{V} \opf{G}_0 \opf{T}$ accounts for repeated scattering events with the obstacles via $\opf{G} = \opf{G}_0 + \opf{G}_0 \opf{T} \opf{G}_0$.
If $\opf{t}_i$ solves the scattering problem for a single impurity $\opf{v}(\vec{s}_i)$ then $\opf{T}$ can be formally expressed by a multiple-scattering expansion~\cite{BallentineBook:2003}
\begin{align}
 \opf{T} = \sum_{i=1}^{N_\text{i}} \opf{t}_i + \sum_{\substack{j,k = 1 \\ j \neq k}}^{N_\text{i}} \opf{t}_j \opf{G}_0 \opf{t}_k +\!\! \sum_{\substack{l,m,n = 1 \\ l \neq m ,\, m \neq n}}^{N_\text{i}}\!\! \opf{t}_l \opf{G}_0 \opf{t}_m \opf{G}_0 \opf{t}_n + \dotsb .
\end{align}
The first sum describes the result of repeated scattering with a single impurity; the second encodes additional subsequent interactions with a different impurity, etc.
After disorder average $\disorder{\opf{T}}$, only the first term contributes to order $\mathcal{O}(n)$ in the obstacle density $n$, while the remaining contributions involve at least pairs of obstacles.
Since disorder averaging restores translational invariance, we obtain $\matrixbraket{\vec{k}}{\disorder{\opf{T}}}{\vec{k}} = n N \maf{t}(\vec{k}) + \mathcal{O}(n^2)$ as the only nonvanishing contribution with the forward-scattering amplitude $\maf{t}(\vec{k}) = \matrixbraket{\vec{k}}{\opf{t}}{\vec{k}}$ of a single impurity located, say, at the origin.
The corresponding Green function is then obtained as
\begin{align}
\disorder{\maf{G}}(\vec{k}) = \maf{G}_0(\vec{k}) + n N \maf{t}(\vec{k}) \maf{G}_0(\vec{k})^2 + \mathcal{O}(n^2),
\end{align}
and comparison with Eq. \eqref{eq:self_energy_opf} reveals $\Sigma(\vec{k}) = nN\maf{t}(\vec{k}) + \mathcal{O}(n^2)$.
Since every single scatterer affects the propagator to order $\mathcal{O}(1/N)$ the product $N \maf{t}(\vec{k})$ converges to a finite value in the limit of large lattices $N \to \infty$.

The remaining task is to determine the matrix elements $\maf{t}(\vec{k})$.
By nearest-neighbor hopping, the operator identity $\opf{t} = \opf{v} + \opf{v}\opf{G}_0 \opf{t} = \opf{v} + \opf{t}\opf{G}_0 \opf{v}$ reveals that the only nonvanishing elements of $\matrixbraket{\vec{r}}{\opf{t}}{\vec{r}'}$ in the real-space basis correspond to  $\vec{r}, \vec{r}' \in \{\vec{0}, \pm \vec{e}_x, \pm \vec{e}_y\}$, i.e., the origin and its neighbors.
Hence, the problem reduces to inverting of a $5 \times 5$ matrix, the solution being expressed in terms of the real-space matrix elements of the unperturbed Green function
\begin{align} \label{eq:propagators}
 \matrixbraket{\vec{r}}{\opf{G}_0}{\vec{r}'} = \int_\text{BZ} \frac{\diff^2 k}{(2\pi)^2} \frac{\exp[\img \scalar{\vec{k}}{(\vec{r} - \vec{r}')}]}{E - \epsilon(\vec{k})},
\end{align}
where the integral extends over the first Brillouin zone $\text{BZ} = \{(k_x, k_y) : k_x, k_y \in [-\pi, \pi[\, \}$.
The required integrals can be calculated analytically in terms of elliptic integrals.
Explicit results for $\disorder{\maf{G}}(\vec{k})$ can be generated in principle by using computer algebra to handle the matrix inversion as shown in the Supplemental Material below.
In the time domain the different moments can be obtained by a suitable derivative of the moment-generating function $F(\vec{k},t)$ with respect to the wave number $\vec{k}$.
In particular, the time-dependent velocity response is obtained from $\img \partial^2 F(\vec{k},t)/\partial t \partial k_x|_{\vec{k} = 0}$, equivalently, in the frequency domain, the corresponding expression reads $\img E \partial \disorder{\maf{G}}(\vec{k})/\partial k_x|_{\vec{k} = 0}$.

\paragraph{Discussion.---}
We have implemented stochastic simulations (see the Supplemental Material below) of the model for tracers starting at an impurity-free site to explore the range of validity  of the low-density expansion for the velocity response.
To make connection to theory, we have to correct for the fraction $n$ of immobile random walkers by normalizing with $1/(1-n) = 1 + n + \mathcal{O}(n^2)$ and retain only terms to first order.
Explicitly, the velocity response in the frequency domain then assumes the form
\begin{align} \label{eq:velocity_E} 
v(E) = \frac{v_0}{E} + n\frac{v_0}{E} + n\frac{\img}{E} \frac{N\partial t(\vec{k})}{\partial k_x} \bigg|_{\vec{k} = 0},
\end{align}
where $v_0 = \img \partial \epsilon(\vec{k})/\partial k_x |_{\vec{k} = 0} = \tanh(F/4)$ is the drift velocity on the impurity-free lattice.

\begin{figure}[t]
\centering
\includegraphics[scale=0.8]{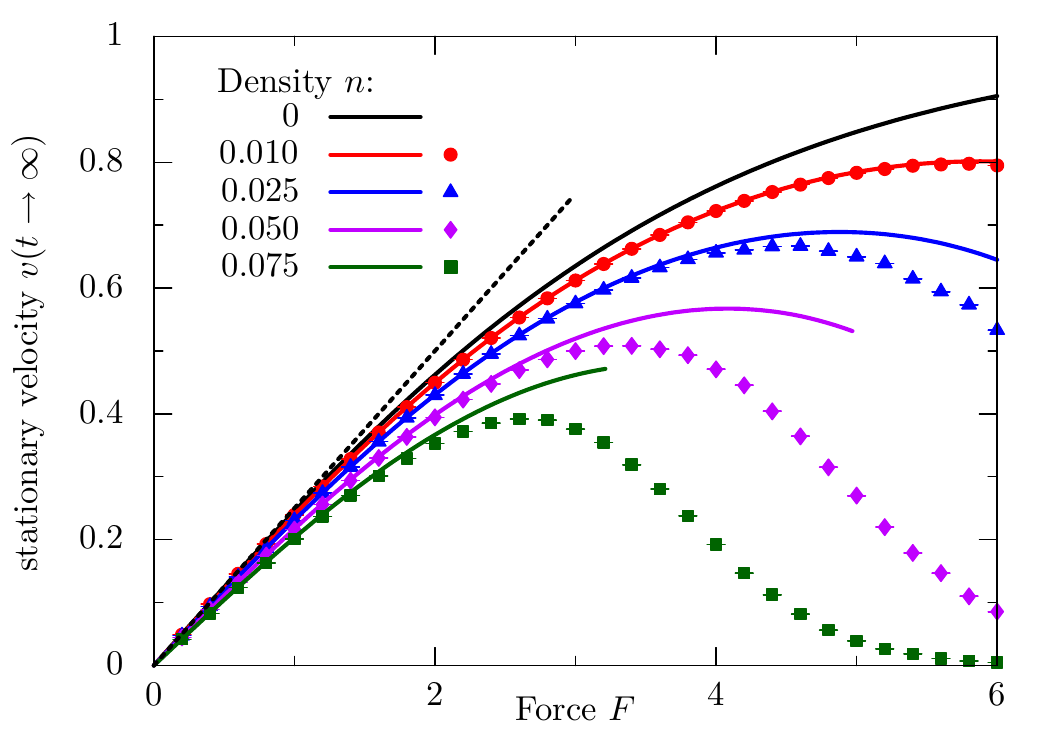}
\caption{
Obstacle-induced velocity response in the stationary state $v(t \to \infty)$ as a function of the force $F$ for increasing obstacle density $n$.
Symbols (with error bars) correspond to stochastic simulations and solid lines represent the first-order approximation in $n$.
The dashed line indicates the linear response approximation in first order of $n$.
}
\label{fig:delta_v}
\end{figure}

The immediate response to the step force is determined by the mean velocity of the first jump event $v(t \to 0) = (1-n) v_0$, where the factor $(1-n)$ accounts for the fraction of accessible sites.
This result can also be inferred from the high-frequency behavior $v(t \to 0) = \lim_{E \to \infty} E v(E)$. 
As time progresses correlations build up since the tracer is scattered off an obstacle many times before it encounters the next impurity.
For long times a stationary state is reached where the probability density of the tracer accumulates in front of the obstacles.
The stationary velocity is encoded in the low-frequency behavior $v(t \to \infty) = \lim_{E \to 0} E v(E)$ and explicit expressions are deferred to the Supplemental Material below. 
The analytic result for the stationary limit $v(t \to \infty)$ is compared to stochastic simulations in Fig.~\ref{fig:delta_v} for different obstacle densities over a wide range of biases.
The simulation data nicely corroborate the theory; deviations become apparent for large forces as the obstacle density increases.
Thus the range of validity of the low-density approximation depends on the driving and the fraction of blocked sites.
Note that small forces correspond to $F \ll 1$ whereas at the strong bias $F = 6$ the tracer already hops $e^6 \approx 400$ times more often in the direction of the field rather than against it.
Interestingly, the stationary velocity displays a maximum (see also the Supplemental Material below), since for large forces the particle runs most of the time against the obstacle rather than around it.
The resulting nonmonotonic behavior is already captured by the theory, yet the suppression of the stationary velocity is underestimated at strong forces indicating that contributions of higher order in the density become relevant.
For small driving the velocity response can be expanded in the force
\begin{align} \label{eq:nonanalytic_expansion}
v(t \to \infty) = D F + \frac{n}{16}\Big(\frac{\pi}{4} - 1\Big) F^3 \ln(F) + \mathcal{O}(F^3) . 
\end{align} 
The leading term corresponds to the linear response consistent with the long-time diffusion coefficient $D = [1 - (\pi - 1)n]/4$ in first order in $n$ as obtained from the VACF $Z(t)$ in equilibrium~\cite{Nieuwenhuizen:1986}.
Already the leading correction is nonanalytic in $F$ and the singularities originate from poles in the integrals for the propagators in the long-wavelength limit [Eq. \eqref{eq:propagators}]~(see the Supplemental Material below). 

The preceding result fits well into the context of nonanalytic dependences of dynamic quantities~\cite{EvansMorrissBook:2008}.
For sheared systems, these contributions have been predicted in Ref.~\cite{KawasakiGunton:1973} and rigorously demonstrated by methods of kinetic theory~\cite{ErnstCichocki:1978}.
For the driven thermostatted Lorentz gas in continuum, nonanalytic contributions in the Lyapunov exponents and the collision frequency have been derived analytically~\cite{PanjaDorfman:2000} and even confirmed by simulations~\cite{DellagovanBeijeren:2001}.

\begin{figure}[t]
\centering
\includegraphics[scale=0.8]{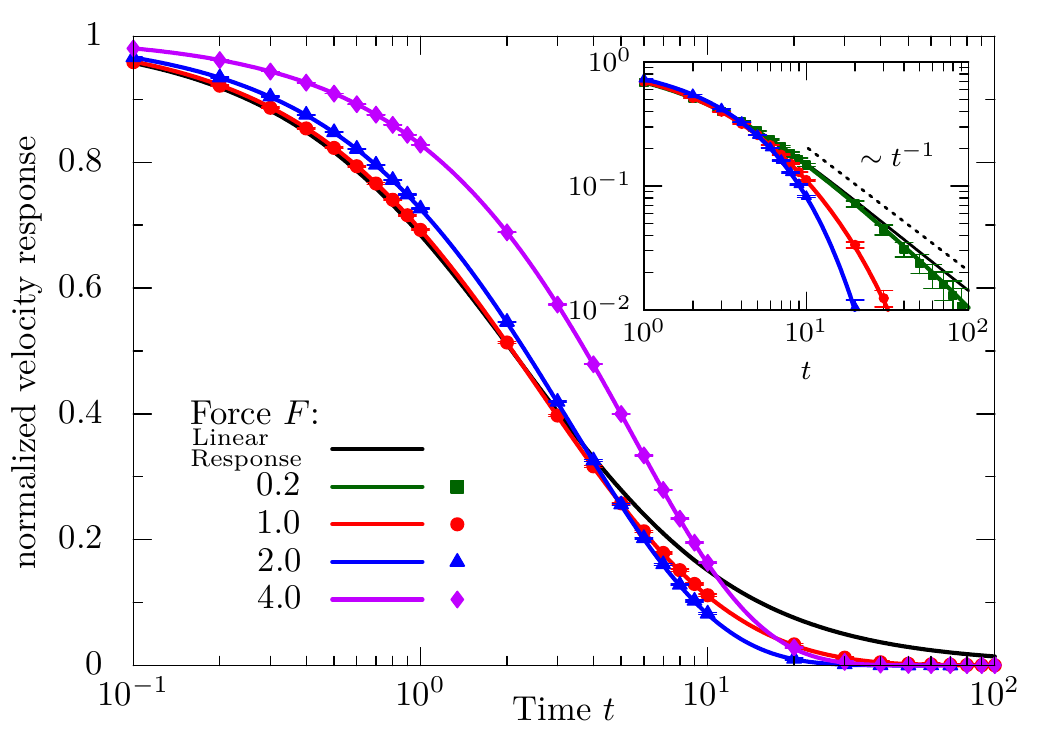}
\caption{
Time-dependent approach to the stationary velocity $v(t) - v(t \to \infty)$ normalized to its initial value.
Symbols (with error bars) correspond to simulation results for increasing force $F$ at density $n = 0.001$.
Solid lines represent the theoretical prediction.
Inset: same quantity in double logarithmic representation.
The dashed line indicates the power-law decay $\sim t^{-1}$ predicted by linear response.
}
\label{fig:velocity_response}
\end{figure}

The time-dependent response $v(t)$ is encoded in the frequency dependence of $v(E)$ [Eq.~\eqref{eq:velocity_E}].
The Fourier back transform is achieved numerically for purely imaginary $E$ and compared to stochastic simulations in Fig.~\ref{fig:velocity_response} (see the Supplemental Material below for details).
The short-time response is more and more delayed as the force increases, whereas the long-time behavior depends nonmonotonically on $F$.
As expected, the velocities follow the linear response result for longer and longer times as the driving becomes weaker.
However, upon closer inspection the power-law tail $\sim t^{-1}$ no longer determines the asymptotic long-time behavior for any finite bias $F>0$; rather the decay is exponentially fast.
This observation can be rationalized by the fact that the propagators [Eq. \eqref{eq:propagators}] become analytic functions in the vicinity of $E = 0$ for any nonzero $F$. 
Hence there is a nonzero radius of convergence $\mathcal{O}(F^2)$ in the complex $E$ plane, which determines the slowest decay rate.
More precisely, the long-time decay can be elaborated to $\sim t^{-1}\exp(-F^2t/16)$ for $F \to 0$ (see the Supplemental Material below).

\paragraph{Conclusion.---}
The stationary and transient velocity of a tracer particle in response to an arbitrarily strong step force on a disordered lattice has been calculated exactly to first order in the obstacle density.
One of our main results is that the persistent correlations expected from the fluctuation-dissipation theorem are decorated by an exponentially fast decay emerging in the nonlinear regime.
Thus, the long-time tail is fragile and linear response breaks down even for arbitrarily small forces.
In that sense, van Kampen's objection, that the chaotic trajectories render linear response invalid, is correct as far as persistent memory is concerned.

The mechanism that leads to the suppression of persistent correlations is transferable also to higher dimension (see the Supplemental Material below) and should generically apply to all equilibrium correlation functions displaying a long-time tail.
The argument is that they all originate from long-wavelength singularities in the unperturbed propagators, yet at finite driving these poles acquire a finite distance from the origin restoring analytic behavior, at least if the driving couples to the relevant conserved quantities.

For the stochastic Lorentz gas, the hydrodynamic mode is single-particle diffusion and the tail originates from the algebraic decay of the return probability $\sim t^{-d/2}$ for long times~\cite{vanBeijeren:1982}, which in the driven case is decorated by an exponential decay in second order in the force, pulling the particle away from the impurity.

For weak driving a power series in the force introduces higher order response functions for which formal expressions in terms of equilibrium correlation functions are available~\cite{Morita:1986, Diezemann:2012, Andrieux:2007}. 
Usually these are difficult to evaluate analytically; however, our calculation includes the response functions to arbitrarily high order for finite times.
For infinite times, these finite order response functions diverge but can be properly resummed to yield the logarithms in the expansion of the terminal velocity for weak driving.
Thus, the assumption that the steady state displays analytic behavior for small driving is in general unjustified.
 
Let us emphasize that the linear response theory in a nonequilibrium state~\cite{BaiesiMaes:2009, Seifert:2012} is complementary to our approach since they discuss a generalization of the Kubo formula for small perturbation around a nonequilibrium state. 
Our calculation includes as a special case of their formalism the incremental change of the steady velocity upon increasing the force by an infinitesimal amount. 

\begin{acknowledgments}
We thank H. van Beijeren for discussions, in particular on the conditions under which the long-time tails are fragile.
We gratefully acknowledge support by the DFG research unit FOR1394, ``Nonlinear response to probe vitrification.''
\end{acknowledgments}

\onecolumngrid

\section{Supplemental Material}

\subsection{Unperturbed propagators} 
In the time domain, the unperturbed Green functions $\matrixbraket{\vec{r}}{\opf{G}_0}{\vec{r}'}$ [Eq.~\eqref{eq:propagators}] are expressed by 
\begin{align} \label{eq:propagators_time}
\matrixbraket{\vec{r}}{\opf{U}_0(t)}{\vec{r}'} = \int_{\text{BZ}} \frac{\diff^2 k}{(2 \pi)^2} \exp[\img \scalar{\vec{k}}{(\vec{r}-\vec{r}')}] \exp[\epsilon(\vec{k})t] .
\end{align}
With the bare diffusion coefficient $2 D_y = 1/[\cosh(F/2) + 1]$ in $y$-direction and the drift velocity $v_0 = 2 D_y \sinh(F/2) = \tanh(F/4)$, 
the eigenvalue $\epsilon(\vec{k}) = \epsilon_x(k_x) + \epsilon_y(k_y)$ can be decomposed in contributions longitudinal $\epsilon_x(k_x) = - 2 D_y \cosh(F/2) [1 - \cos(k_x)] - \img v_0 \sin(k_x)$ and transversal $\epsilon_y(k_y) = - 2 D_y [1 - \cos(k_y)]$ to the direction of the force $F \vec{e}_x$.
Hence, the longitudinal and transversal motion are independent such that the integrals in Eq. \eqref{eq:propagators_time} decouple.
Both can be evaluated by the formula~\cite[Eq. (64)]{MontrollScher:1973}
\begin{align}
\int_{-\pi}^{\pi} \frac{\diff k}{2 \pi} \exp[\img k x] \exp[\alpha \cos(k) - \img \beta \sin(k)] = \Bigg[\frac{\alpha + \beta}{\sqrt{\alpha^2 - \beta^2}}\Bigg]^x \text{I}_x \big(\sqrt{\alpha^2 - \beta^2}\big) \end{align}
which can be derived by the contour integral representation of the modified Bessel function of the first kind of integer order $x$, $\text{I}_x(\dotso)$.
Then, the matrix elements of the time-evolution operator assume the form
\begin{align} \label{eq:U_factorized}
\matrixbraket{\vec{r}}{\opf{U}_0(t)}{\vec{r}'} = \exp\bigg[\frac{F}{2} (x-x')\bigg] \matrixbraket{\vec{r}}{\opf{u}(t)}{\vec{r}'}
\end{align}
where the exponential accounts for the asymmetry introduced by the bias, whereas the remaining time evolution $\matrixbraket{\vec{r}}{\opf{u}(t)}{\vec{r}'} = \exp(-t) \text{I}_{x-x'}(2 D_y t) \text{I}_{y - y'}(2 D_y t)$ coincides with the free motion except for the change of the diffusion coefficient $D_y|_{F = 0}$ to $D_y$. 
This modification translates in the frequency domain to 
\begin{align}
\int_0^\infty \diff t \exp(-Et) \matrixbraket{\vec{r}}{\opf{u}(t)}{\vec{r}'} = \frac{1}{4 D_y} \int_{\text{BZ}} \frac{\diff^2 k}{(2 \pi)^2} \frac{\exp[\img \scalar{\vec{k}}{(\vec{r}-\vec{r}')}]}{E' - \epsilon(\vec{k})|_{F = 0}} .
\end{align}
with $E' = (1 + E)/4 D_y - 1$.
Hence, the pole at $E = 0$ for $F = 0$ in the long-wavelength limit is shifted to $E = 4 D_y - 1 = - \tanh^2(F/4) = -F^2/16 + \mathcal{O}(F^4)$.

The leading singularity at $E' = 0$ in the propagator for $\vec{r} = \vec{r}'$ can be elaborated to $\sim -\ln(E')/\pi$ for dimension $d = 2$, $\sim -(3/\pi) \sqrt{3E'/2}$ for $d = 3$, and $\sim (4/\pi^2) E' \ln(E')$ for $d = 4$ via dimensional regularization~\cite{Leibbrandt:1975}.
These correspond to the long-time tails $\sim (2 \pi t/d)^{-d/2}$ of $\matrixbraket{\vec{r}}{\opf{u}(t)}{\vec{r}'}|_{F = 0}$.

For $F \to 0$, the shift of the pole can be attributed to an exponential decoration of the long-time tail via 
\begin{align}
\int_0^\infty \diff t \exp(-Et)\exp(-F^2t/16) f(t) \approx f(E + (1+E)F^2/16)
\end{align}
for $E \to 0$.
We have checked that our numerical solution indeed reflects this property.

\subsection{Calculation of the single obstacle t-matrix} 
The singe obstacle $t$-matrix encodes multiple scattering events of the tracer with a single impurity $\opf{v}$ and fulfills the operator identity $\opf{t} = \opf{v} + \opf{v}\opf{G}_0\opf{t} = \opf{v} + \opf{t}\opf{G}_0\opf{v}$.
In our case of nearest-neighbor hopping dynamics, the matrix elements $\matrixbraket{\vec{r}}{\opf{v}}{\vec{r}'}$ are nonvanishing only if the impurity or one of its neighboring sites is involved.

Thus the operator identity can be read as a $5 \times 5$ matrix inversion problem.
Consequently, we can restrict our calculations to the subspace spanned by the relevant sites around the impurity and order the rows and columns for the matrix forms of the operators in the real-space basis via the scheme 
\begin{align*}
\begin{matrix}
  & 1 &   \\
2 & 3 & 4 \\
  & 5 &   \\
\end{matrix} 
\end{align*}
where the impurity is located at the origin $\vec{0}$ numbered by $3$.
Then, the operator $\opf{v}$ in the subspace assumes the form
\begin{align}
\maf{v} = D_y
\begin{pmatrix}
1 & 0 & -1 & 0 & 0 \\
0 & e^{F/2} & -e^{-F/2} & 0 & 0 \\
-1 & -e^{F/2} & 1/D_y & -e^{-F/2} & -1 \\
0 & 0 & -e^{F/2} & e^{-F/2} & 0 \\
0 & 0 & -1 & 0 & 1
\end{pmatrix} .
\end{align}
For the matrix form of $\opf{G}_0$, we use Eq.~\eqref{eq:U_factorized} for simplifications and define the Laplace transform of the time-dependent part by $\matrixbraket{\vec{r}}{\opf{g}}{\vec{r}'} = \int_0^\infty \diff t \exp(-Et) \matrixbraket{\vec{r}}{\opf{u}(t)}{\vec{r}'}$.
Hence, we obtain
\begin{align}
\maf{G}_0 = \begin{pmatrix}
\maf{g}_{33} & e^{F/2} \maf{g}_{12} & \maf{g}_{13} & e^{-F/2} \maf{g}_{12} & \maf{g}_{15} \\
e^{-F/2} \maf{g}_{12} & \maf{g}_{33} & e^{-F/2} \maf{g}_{13} & e^{-F} \maf{g}_{15} & e^{-F/2} \maf{g}_{12} \\
\maf{g}_{13} & e^{F/2} \maf{g}_{13} & \maf{g}_{33} & e^{-F/2} \maf{g}_{13} & \maf{g}_{13} \\
 e^{F/2} \maf{g}_{12} & e^{F} \maf{g}_{15} & e^{F/2} \maf{g}_{13} & \maf{g}_{33} & e^{F/2} \maf{g}_{12} \\
\maf{g}_{15} & e^{F/2} \maf{g}_{12} & \maf{g}_{13} & e^{-F/2} \maf{g}_{12} & \maf{g}_{33} 
\end{pmatrix}
\end{align}
where we abbreviated $\matrixbraket{\vec{0}}{\opf{g}}{\vec{0}} = \maf{g}_{33}$, $\matrixbraket{\vec{e}_y}{\opf{g}}{-\vec{e}_x} = \maf{g}_{12}$, $\matrixbraket{\vec{e}_y}{\opf{g}}{\vec{0}} = \maf{g}_{13}$, and $\matrixbraket{\vec{e}_y}{\opf{g}}{-\vec{e}_y} = \maf{g}_{15}$.
Then, with computer algebra, it is straightforward to calculate $(1 - \maf{G}_0\maf{v})^{-1}$ in the distinguished subspace. Using the identities
\begin{equation} 
\begin{split} \label{eq:free_motion}
\matrixbraket{\vec{0}}{(E - \opf{H}_0)\opf{G}_0}{\vec{0}} = 1 \quad &\Longleftrightarrow \quad   (1 + E) \maf{g}_{33} - 4 D_y \maf{g}_{13} = 1 \\
\matrixbraket{\vec{0}}{(E - \opf{H}_0)\opf{G}_0}{\vec{e}_y} = 0 \quad &\Longleftrightarrow \quad   (1 + E) \maf{g}_{13} - D_y (2 \maf{g}_{12} + \maf{g}_{15} + \maf{g}_{33}) = 0 
\end{split}
\end{equation}
leads to further simplifications. 
Then, transforming to the plane-wave basis with the formula $N \maf{t}(\vec{k}) = N\matrixbraket{\vec{k}}{\opf{t}}{\vec{k}} = \sum_{\vec{r}, \vec{r}'} \exp[-\img \scalar{\vec{k}}{(\vec{r} - \vec{r}')}] \matrixbraket{\vec{r}}{\opf{v}(1 - \opf{G}_0 \opf{v})^{-1}}{\vec{r}'}$ completely determines the self-energy to first order in the density $\Sigma(\vec{k}) = n N \maf{t}(\vec{k}) + \mathcal{O}(n^2)$.
The explicit expression is too long to be presented here.
Rather we focus on the relevant contribution for the velocity response [Eq.~\eqref{eq:velocity_E}], which can be calculated by computer algebra and brought into the form
\begin{align} \label{eq:derivative_t}
\img N \frac{\partial \maf{t}(\vec{k})}{\partial k_x}\bigg|_{\vec{k} = 0} = \frac{g_{N1} \sinh(F/2)}{ g_{D1} + g_{D2} \cosh(F/2)}
\end{align}
with 
\begin{equation}
g_{N1} = D_y [2 (1 - 4D_y) \maf{g}_{13} + E \maf{g}_{33}](2 \maf{g}_{12} - \maf{g}_{15} - \maf{g}_{33}) + 2D_y(\maf{g}_{15}-\maf{g}_{33}) - E \maf{g}_{13} ,
\end{equation}
\begin{equation}
\begin{split}
g_{D1} =\ &\maf{g}_{33} + D_y [1 + D_y (\maf{g}_{15}-\maf{g}_{33})] [2 \maf{g}_{13}^2-\maf{g}_{33} (\maf{g}_{15}+\maf{g}_{33})] \\
&+D_y^3 (2 \maf{g}_{12}-\maf{g}_{15}-\maf{g}_{33}) (\maf{g}_{15}-\maf{g}_{33}) [4 \maf{g}_{13}^2-\maf{g}_{33} (2 \maf{g}_{12}+\maf{g}_{15}+\maf{g}_{33})],
\end{split}
\end{equation}
and
\begin{equation}
\begin{split}
g_{D2} = 2 D_y \maf{g}_{13}^2 [1+D_y  (4 \maf{g}_{12}- \maf{g}_{15}-3 \maf{g}_{33})] - 2 D_y \maf{g}_{33} \{D_y  [2 \maf{g}_{12}^2-\maf{g}_{33} (\maf{g}_{15}+\maf{g}_{33})]+\maf{g}_{33}\} .
\end{split}
\end{equation}

The remaining task is to determine explicit expressions for the four propagators $\maf{g}_{33}$, $\maf{g}_{12}$, $\maf{g}_{13}$, and $\maf{g}_{15}$.
They can be expressed in terms of complete elliptic integrals $\mathbf{K}[k] = \int_0^{\pi/2} \diff \theta /\sqrt{1 - k \sin^2(\theta)}$ and $\mathbf{E}[k] = \int_0^{\pi/2} \diff \theta \sqrt{1 - k \sin^2(\theta)}$.
Starting with $\maf{g}_{12}$, the necessary integral $\maf{g}_{12} = \int_0^\infty \diff t \exp[-(1 + E) t] \text{I}_1(2 D_y t) \text{I}_1(2 D_y t)$ is equivalent to~\cite[p. 696, Eq. 6.612.5]{Gradshteyn:2007} and with the relations $\mathbf{K}[k] = \mathbf{K}[k/(k-1)]/\sqrt{1-k}$ and $\mathbf{E}[k] = \sqrt{1 - k} \,\mathbf{E}[k/(k-1)]$, one obtains 
\begin{align}
\maf{g}_{12} = -\frac{(1+E)^2 \mathbf{E}[(4 D_y / (1 + E))^2]}{4 D_y^2 \pi (1+E)} - \frac{(8 D_y^2 - (1 + E)^2)\mathbf{K}[(4 D_y / (1 + E))^2]}{4 D_y^2 \pi (1 + E)}.
\end{align}
Continuing with $\maf{g}_{33}$, the integral $\maf{g}_{33} = \int_0^\infty \diff t \exp[-(1 + E) t] \text{I}_0(2 D_y t) \text{I}_0(2 D_y t)$ evaluates to
\begin{align}
\maf{g}_{33} = \frac{2 \mathbf{K}[(4 D_y/(1 + E))^2]}{\pi (1 + E)}
\end{align}
with the formula~\cite[p. 696, Eq. 6.612.4]{Gradshteyn:2007}.
Then, the remaining two propagators follow immediately by Eq.~\eqref{eq:free_motion}.

\subsection{Velocity response}
In first order of the force $F$, the velocity response determined by Eq.~\eqref{eq:velocity_E} reduces with computer algebra to
\begin{align}
v(E) = F \frac{Z(E)}{E} + \mathcal{O}(F^3),
\end{align}
where
\begin{align}
Z(E) = \frac{1}{4} - \frac{n}{4}\bigg[-\frac{2 \pi}{\pi E - 2 (1 + E) \mathbf{E}[(1+E)^{-2}]} - 1\bigg]
\end{align}
represents the velocity-autocorrelation function (VACF) in equilibrium~\cite{Nieuwenhuizen:1986}.
Thus, for small forces, the velocity in the stationary state is described by $v(t \to \infty) = \lim_{E \to 0} Ev(E) = D F$ with $D = Z(0) = [1 - n (\pi - 1)]/4$.
For arbitrary strong driving, the stationary velocity is calculated with computer algebra and assumes the form 
\begin{align}
v(t \to \infty) = v_0 - n v_0 \bigg[\frac{(4 D_y)^{-1} (\pi - 2 \text{E})}{\text{E} - [1 - (4D_y)^2]\text{K}} - \frac{[1 - (4 D_y)^{-1}](\pi - 2\text{E})}{2 D_y \pi - \text{E} + (1 - 4 D_y)\text{K}} + 1\bigg]
\end{align}
where we abbreviated $\text{K} = \mathbf{K}[(4 D_y)^2]$, and $\text{E} = \mathbf{E}[(4 D_y)^2]$. 

\subsection{Nonmonotonic behavior in the terminal velocity}
\begin{figure}
\includegraphics[scale=0.8]{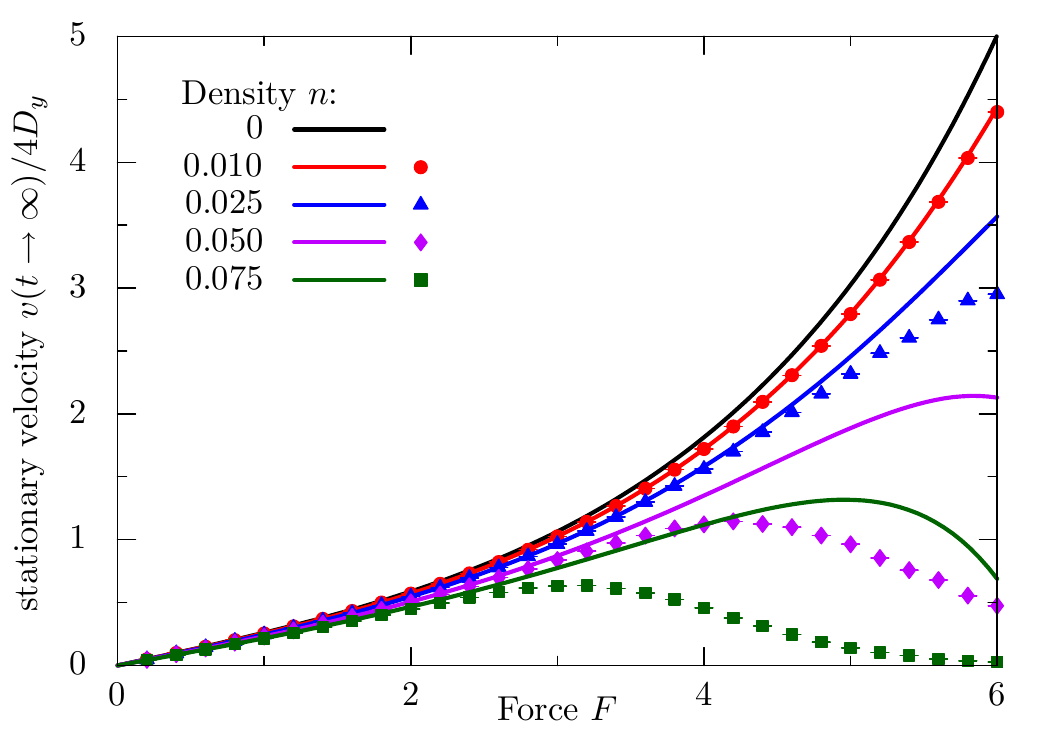}
\caption{
Obstacle-induced velocity response in the stationary state $v(t \to \infty)/4 D_y$ for unnormalized transition rates as a function of the force $F$ for increasing obstacle density $n$.
Symbols (with error bars) correspond to stochastic simulations and solid lines represent the first-order approximation in $n$.
}
\label{fig:unnorm_stat_vel}
\end{figure}

In the definition of the model, we used normalized transition rates.
However, results for unnormalized rates can be directly obtained by a suitable transformation.
Multiplying the stationary velocity $v(t \to \infty)$ by the factor $(4 D_y)^{-1} = (e^{F/2} + e^{-F/2} + 2)/4$, one obtains the steady state velocity for the transition rates $W(\pm \vec{e}_x)/4 D_y = e^{\pm {F/2}}/4$ and $W(\pm \vec{e}_y)/4 D_y = 1/4$, shown in Fig.~\ref{fig:unnorm_stat_vel}.
For these rates, the nonmonotonic behavior of the terminal velocity is still present and correspondingly not specific to the normalization used in the main model.

\subsection{Numerical inversion}
For a real function $f(t)$ we employ the one-sided Fourier transform 
\begin{align}
f(E) = \int_0^\infty \diff t \exp(-E t) f(t)
\end{align}
with $E = \sigma + \img \omega$ in the complex right half plane.
Thus, $\Real[f(E)]$ is twice the Cosine transform of $f(|t|) \exp(-\sigma |t|)$ and one finds by inversion,
\begin{align} \label{eq:fourier_inversion}
f(t) = \frac{2 e^{\sigma t}}{\pi} \int_0^\infty \diff \omega \Real[f(\sigma + \img \omega)]\cos(\omega t), \quad t \geq 0.
\end{align}
Due to the long-time limit $v(t \to \infty)$, $v(E)$ acquires a simple pole at $E = 0$ and we are not allowed to set $\sigma = 0$ for the back-transform.
Hence, we apply Eq.~\eqref{eq:fourier_inversion} with $\sigma = 0$ to $v(E) - v(t \to \infty)/E$ and obtain the change in the velocity $v(t) - v(t \to \infty)$. 
Numerically, the back-transform is achieved by a suitable Filon formula~\cite{Tuck:1967} and multiplication by $[v(t \to 0) - v(t \to \infty)]^{-1}$ leads to the normalized velocity response shown in Fig.~\ref{fig:velocity_response}.

\subsection{Stochastic simulation}
\begin{figure}[h]
\centering
\includegraphics[scale=0.6]{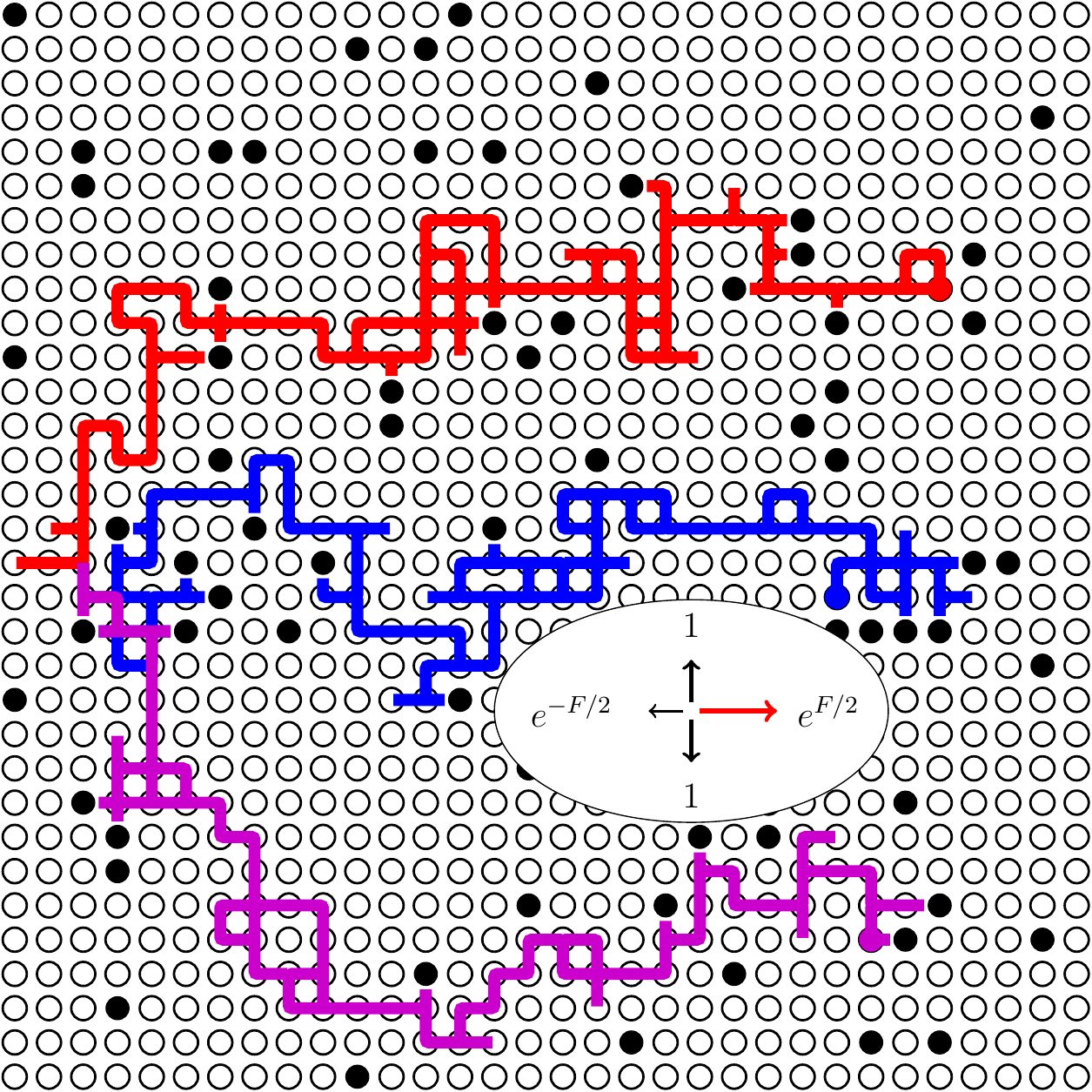}
\caption{Sample trajectories of a random walker in a typical obstacle realization. The force pulls the tracer to the right resulting in a net drift. At low densities, the dominant contribution is a series of uncorrelated collision events with isolated obstacles.}
\label{fig:lattice}
\end{figure}

The simulation data were generated by random walks in discrete time on lattices of size $1024 \times 1024$ with periodic boundary conditions, illustrated in Fig.~\ref{fig:lattice}. 
For each obstacle density $n$, $10^3$ different configurations of the disorder were realized and for each realization we simulated over $10^7$ trajectories consisting of $150$ jumps of the tracer. 
From the mean distance traveled along the field after $k$ jumps, $\langle \Delta x \rangle(k)$, the simulated velocity response was calculated by the formula
\begin{align}
v(t) = \frac{\diff}{\diff t} \sum_{k = 0}^\infty \frac{t^k}{k!} e^{-t} \langle \Delta x \rangle(k) = \sum_{k = 0}^\infty \frac{t^k}{k!} e^{-t} [\langle \Delta x \rangle(k + 1) - \langle \Delta x \rangle(k)] .
\end{align}
Moreover, the necessary number of trajectories were reduced by a method used in~\cite{Frenkel:1987}, where only differences to the trajectory in absence of the obstacles are relevant.
In order to obtain the data for the stationary velocity in Fig.~\ref{fig:delta_v}, the starting point of the tracers for the new trajectory coincided with the last position in the preceding trajectory. 
For the time-dependent velocity response in Fig.~\ref{fig:velocity_response}, the random walker was placed on a randomly chosen accessible site after the trajectory was completed.

\bibliographystyle{apsrev4-1}
\end{document}